\documentclass[11pt]{article}

\usepackage{latexsym,amssymb,amsfonts,graphicx,makeidx,a4wide,theorem,rotating,ifthen,amsmath,subfigure,epsfig}

\usepackage{amsfonts,epsfig,amssymb}
\newcommand{\graph} {\mbox{val}}
\newcommand{\val} {\mbox{val}}
\newcommand{\CW} {\mbox{CW}}

\newcommand{\lab} {\mbox{lab}}

\newcommand{\MSOA}{\mbox{MSO}_1}
\newcommand{\pnp}{\mbox{P}^{\mbox{NP}}_{\parallel}}

\newtheorem{theorem}{Theorem}[section]

\newtheorem{definition}[theorem]{Definition}

\newtheorem{problem}[theorem]{Problem}

\newenvironment{proof}{\noindent{\bf Proof~}}{\null\hfill $\Box$\par\medskip}

\begin{document}

\title{Polynomial algorithms for protein similarity search for restricted mRNA structures}

\author{Frank Gurski\thanks{Heinrich-Heine Universit\"at D\"usseldorf,
Department of Computer Science, D-40225 D\"usseldorf, Germany,
E-Mail: gurski-corr@acs.uni-duesseldorf.de, }}

\bigskip

\maketitle

\begin{abstract}
In this paper we consider the problem of computing an mRNA sequence of maximal similarity for a given mRNA 
of secondary structure constraints, introduced by Backofen et al. in \cite{BNS02} denoted as the MRSO problem.  The problem is known to be NP-complete for planar associated implied structure graphs of vertex degree at most 3. In \cite{BFHV05} a first polynomial dynamic programming algorithms for MRSO on implied structure graphs with maximum vertex degree 3 of bounded cut-width is shown. 
We give a simple but more general polynomial dynamic programming solution for the MRSO
problem for associated implied structure graphs of bounded clique-width. Our 
result implies that MRSO is polynomial for graphs of bounded
tree-width, co-graphs, $P_4$-sparse graphs, and distance hereditary graphs. 
Further we conclude that the problem of comparing two solutions for MRSO
is hard for the class $\pnp$, which is defined as the set of
problems which can be solved in polynomial time with a number of parallel
queries to an oracle in NP. 

\end{abstract}


\section{Introduction}


One of the main processes in biology is the transformation of DNA into proteins. 
This process is divided into two steps. The fist step is the transcription, which
copies the DNA into a certain RNA molecule called messenger RNA (mRNA). Each
mRNA is a string of four types of nucleotides, i.e. elements of $\{A,C,G,U\}$. $(A,U)$ and $(C,G)$
are known as the complementary nucleotide pairs. Every string
of three nucleotides is called a codon.
The second step is the translation, which converts block wise a codon 
in the mRNA into an amino acid. Every protein is the result of a translation of some mRNA.

We can represent every mRNA as a graph by considering its nucleotides as vertices and possible  edges (so called bonds) between vertices representing complementary nucleotides. 
The resulting graph is also denoted as (secondary) structure graph of the mRNA. If we consider the codons as vertices we obtain the associated implied structure graph.

In this paper consider the MRna Structure Optimization (MRSO) problem, introduced by Backofen et al. \cite{BNS02,BNS02a}. The problem is to compute an mRNA sequence of maximal similarity for a given mRNA that additionally satisfies some secondary structure constraints. If the input structure graph
of the problem has vertex degree at most one, we will denote the restriction of the problem by MRSO-d1.
The MRSO-d1 problem (and thus also the MRSO problem) has shown to be NP-complete for planar implied structure graphs \cite{BFHV05}. In \cite{BNS02} a linear time algorithm for the MRSO-d1 problem has
been shown for outer-planar implied structure graphs.

A very useful tool to solve hard problems on restricted
inputs is parameterized complexity \cite{DF99}. The main idea is that only input
graphs of a bounded graph parameter $k$ are considered. The running
time of the algorithms is exponential in $k$, but for
fixed $k$ polynomial. Fixed parameter algorithms are frequently used in
computational biology, see e.g. works of Bodlaender et al. \cite{BDFW95,BDFHW95}.
In \cite{BFHV05} first fixed parameter algorithms for the MRSO-d1 problem for implied structure graphs of bounded cut-width are given.
There are several known secondary (tertiary and quaternary) structures which are not simple, recursive, or do not correspond to outer-planar implied structure graphs \cite{Aku00}. Further it seems to be likely that amino acids of more complicated secondary structures will be discovered in the future \cite{BFHV05}.
Therefore, we give a more general fixed parameter solution for MRSO on graphs of bounded clique-width which form a very large class of implied structure graphs.

The clique-width of a graph is defined by a composition mechanism for
vertex-labeled graphs \cite{CO00}. The operations are the vertex disjoint
union, the addition of edges between vertices controlled by a label pair, and
the relabeling of vertices. The clique-width of a graph $G$ is the minimum number of labels needed to define it. Each such composition leads a tree structure. Using this tree structure
a lot of NP-complete graph problems can be solved by dynamic programming in polynomial time for
graphs of bounded clique-width, see e.g. \cite{CMR00,EGW01a,GW06,KR03}.

This paper is organized as follows. In Section 2, we recall the definition of
clique-width and a general method how to solve
graph problems on graph of bounded clique-width. In Section 3, we 
recall the definition of MRSO from Backofen et al. \cite{BNS02,BNS02a}. 
In Section 4, we show a simple but very general polynomial time solution
of the problem for implied structure graphs of bounded clique-width. Our 
result implies that MRSO is polynomial for graphs of bounded
tree-width, co-graphs, $P_4$-sparse graphs, and distance hereditary graphs and re-proofs the
existence of polynomial time algorithms for MRSO-d1 of \cite{BFHV05} for graphs of bounded tree-width and graphs of bounded cut-width. In Section 5,
we briefly conclude that the problem of comparing to solutions for MRSO-d1 
is hard for the class $\pnp$, which is defined as the set of
problems which can be solved in polynomial time with a number of parallel
queries to an oracle in NP.


\section{Clique-width and polynomial time algorithms}


Let $[k]:=\{1,\ldots,k\}$ be the set of all integers between $1$ and $k$.
We work with finite undirected labeled {\em graphs} $G=(V_G,E_G,\lab_G)$,
where $V_G$ is a finite set of {\em vertices} labeled by some mapping
$\lab_G: V_G \to [k]$ and $E_G \subseteq \{ \{u,v\} \mid u,v \in
V_G,~u \not= v\}$ is a finite set of {\em edges}.
The labeled graph consisting of a single vertex labeled by $a \in
[k]$ is denoted by $\bullet_a$. For the definition of special graph classes
we refer  to the survey of Brandst\"adt  et al. \cite{BLS99}.

The notion of clique-width\footnote{This complexity measure was first considered by Courcelle, Engelfriet, and Rozenberg \cite{CER91,CER93}, the notion of  clique-width was introduced 
by Courcelle and Olariu in \cite{CO00}.} for labeled  graphs is defined by Courcelle and Olariu in \cite{CO00}
as follows.

\begin{definition}[Clique-width, \cite{CO00}]
\label{D1}
Let $k$ be some positive integer. The class $\CW_k$ of labeled graphs is
recursively defined as follows.

\begin{enumerate}
\item
The single vertex graph $\bullet_a$ for some $a \in [k]$ is in $\CW_k$.
\item
Let $G,J \in \CW_k$ be two
vertex disjoint  labeled graphs, then $G \oplus J:=(V',E',\lab')$
defined by $V':=V_G  \cup V_J$, $E':=E_G \cup E_J$, and
\[\lab'(u) \ := \ \left\{\begin{array}{ll}
\lab_G(u) & \mbox{if } u \in V_G\\
\lab_J(u) & \mbox{if } u \in V_J\\
\end{array} \right.,\ \forall u \in V'\]
is in $\CW_k$.
\item
Let $a,b \in [k]$ be two distinct integers and $G\in
\CW_k$ be a labeled graph, then
\begin{enumerate}
\item $\rho_{a \rightarrow b}(G):=(V_G,E_G,\lab')$ defined by
\[\lab'(u) \ := \ \left\{\begin{array}{ll}
\lab_G(u) & \mbox{if } \lab_G(u) \not= a\\
        b & \mbox{if } \lab_G(u) = a\\
\end{array} \right.,\ \forall u \in V_G\]
is in $\CW_k$ and
\item
$\eta_{a,b}(G)\ :=\ (V_G,E',\lab_G)$ defined by
$E':=E_G \cup \{ \{u,v\} \mid u,v \in V_G,~u\not=v,~\lab(u)=a,~\lab(v)=b \}$
is in $\CW_k$.
\end{enumerate}
\end{enumerate}
The {\em clique-width} of a labeled graph $G$ is the least integer $k$ such
that $G \in \CW_k$. The {\em clique-width} of an unlabeled graph $G=(V_G,E_G)$ is the smallest integer
$k$, such that there is some mapping  $\lab_G : V_G \to [k]$ such that
the labeled graph $(V_G,E_G,\lab_G)$ has  clique-width at  most $k$.

A class of graphs $\mathcal{L}$ has  {\em bounded clique-width} if there is some integer $k$ such that any graph in $\mathcal {L}$
has clique-width at most $k$, i.e. there is some $k$ such that  ${\mathcal L}\subseteq \CW_k$.
The minimal $k$, if exists, is
defined as  {\em clique-width}
of class ${\mathcal L}$.
\end{definition}

\begin{figure}[ht]
\begin{center}
\begin{picture}(205.5,65)
{
\put(10,5){$G_1$}
\put(20,0){\epsfig{figure=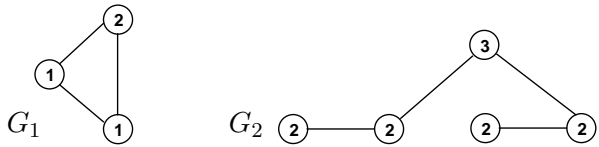,width=7.5cm}}
\put(94,5){$G_2$}}
\end{picture}
\end{center}
\caption{Two labeled graphs $G_1$ and $G_2$ defined by expressions $X_1$ and $X_2$, respectively. The inscriptions in the circles represent the labels of the vertices.}
\label{Fgr}
\end{figure}

An expression built with the operations
$\bullet_a,\oplus,\rho_{a \rightarrow b},\eta_{a,b}$ for integers $a,b \in
[k]$ is called a {\em clique-width $k$-expression}. 
The graph defined by expression $X$ is denoted by $\graph(X)$.
The following two clique-width expressions $X_1$ and $X_2$ define the labeled graphs
$G_1$ and $G_2$ in Fig. \ref{Fgr}.

\[X_1=\eta_{1,2}( (\rho_{2 \to 1}(\eta_{1,2}(\bullet_1 \oplus \bullet_2)))
\oplus \bullet_2)\]

\[X_2= \rho_{1 \to 2}(\eta_{2,3}
(((\eta_{1,2}(\bullet_1 \oplus \bullet_2))  \oplus (\eta_{1,2}(\bullet_1 \oplus \bullet_2)))
\oplus \bullet_3))\]

If a graph $G$ has clique-width at most $k$ then the edge complement
$\overline{G}$ has clique-width at most $2k$ \cite{CO00}. Distance hereditary graphs have clique-width at most 3 \cite{GR00}. 
Co-graphs, i.e. $P_4$-free graphs have clique-width at most 2 \cite{CO00}. Further,
many graph classes defined by a limited number of $P_4$ have bounded
clique-width, e.g. $P_4$-sparse graphs, $P_4$-tidy, and $(q,t)$-graphs \cite{CMR00,MR99}.
The clique-width  of permutation graphs, interval graphs, grids and planar graphs is
not bounded \cite{GR00}. An arbitrary graph with $n$ vertices has clique-width at most $n-r$,
if $2^{r} <  n-r$. 
Every graph of
tree-width at most $k$
has clique-width at most $3\cdot 2^{k-1}$ \cite{CR05}. The recognition problem for graphs of clique-width at most $k$ is still open for $k\ge 4$. Clique-width of
at most $3$ is decidable in polynomial time \cite{CHLRR00}.  Clique-width of at most $2$ is decidable in linear
time \cite{CPS85}. Both algorithms also give a clique-width expression if the input graph
has clique-width at most 3 or clique-width at most 2, respectively.
The clique-width of tree-width bounded graphs is also computable in linear time \cite{EGW03}. Minimizing 
clique-width is NP-complete \cite{FRRS06}.

Courcelle et al. have shown in \cite{CMR00} that all graph properties which are expressible in monadic second order logic with
quantifications over vertices and vertex sets ($\MSOA$-logic) are decidable in
linear time on clique-width bounded graphs.
Furthermore, there are many NP-complete graph problems which are not expressible in
extended $\MSOA$-logic like Hamiltonicity, chromatic number, partition problems, and
bounded degree subgraph problems but which can also be solved in polynomial time on clique-width bounded graphs. The algorithms can be found in  \cite{EGW01a,GW06,KR03}. The proofs are based 
on the following general dynamic programming scheme.

\begin{theorem} [\cite{EGW01a}] \label{T1}
Let $\Pi$ be a graph problem and $k$ be a positive integer.
If there is a mapping $F$ that maps each clique-width $k$-expression $X$
onto some structure $F(X)$,
such that for all clique-width $k$-expressions $X,Y$ and all $a,b\in[k]$
\begin{enumerate}
\item the size of $F(X)$ is polynomially bounded in the size of $X$,
\item the answer to $\Pi$ for $\graph(X)$ is computable in polynomial time from $F(X)$,
\item $F(\bullet_a)$,  is computable in time $O(1)$,
\item $F(X \oplus Y)$ is computable in polynomial time from $F(X)$ and $F(Y)$, and
\item $F(\eta_{a,b}(X))$ and $F(\rho_{a \to b}(X))$, are computable in polynomial time from $F(X)$.
\end{enumerate}
Then for every clique-width $k$-expression $X$, the answer to $\Pi$ for graph $\graph(X)$ is computable in polynomial time from expression $X$. 
\end{theorem}

One of the main important questions is how to find clique-width expressions. For graphs of clique-width at most 3
an expression can be found in polynomial time, as stated above. For graphs of
larger clique-width, approximations of rank-width \cite{OS06,Oum05,Oum06} lead 
approximations of clique-width and a corresponding expression. The best known 
result is the following.

\begin{theorem}[\cite{Oum06}] \label{T2}
For every fixed integer $k$ there is a $O(|V_G|^3)$ algorithm that either outputs a clique-width $(8^{k}-1)$-expression of an input graph $G$, or confirms that the clique-width of $G$ is larger that $k$.
\end{theorem}

\section{MRna Structure Optimization (MRSO)}

In this section we recall the MRna Structure Optimization (MRSO) problem as introduced by Backofen et al. in \cite{BNS02}.

A {\em codon} is a sequence of three {\em nucleotides}, i.e. a string of $\{A,C,G,U\}^3$.
UAA,UAG and UGA are called   {\em stop codons}, the remaining codons represent 20 {\em amino acids}.
An {\em mRNA} is a sequence of $n$ consecutive codons $S=s_1\ldots s_{3n}$ over $\{A,C,G,U\}$, i.e.
each codon of $S$ is of the form $s_{3i-2}s_{3i-1}s_{3i}$ for some $1\leq i\leq n$.

The MRna Structure Optimization (MRSO) problem is defined in \cite{BNS02} as follows. Let
$S=S_1\ldots S_{3n}$  be the nucleotide sequence of an mRNA and let $A=A_1\ldots A_{n}$ be a given amino acid sequence. The problem is to find an approximative mRNA sequence $N=N_1\ldots N_{3n}$ with amino acid sequence $A'=A'_1\ldots A'_{n}$, such
that  $N$ and $S$ have the same secondary structure and $A$ and  $A'$ are of maximum similarity. The similarity between  amino acid sequences is measured by PAM matrices introduced by Dayhoff et al. \cite{DSO78}. We will use $n$ functions
$f_i$, $1\leq i \leq n$ measuring the similarity between $A_i$ and $A'_i$. 

In order to define the MRSO as a general graph problem we  use
the following notions.
Let $\Sigma$ be a finite alphabet (in biological application $\Sigma=\{A,C,G,U\}$
corresponds to the set of nucleotides) and $\Gamma\subseteq \Sigma \times \Sigma$ be a
set of complementary pairs over $\Sigma$ (in biological application $\Gamma=\{(C,G),(A,U)\}$ corresponds to the set of complementary nucleotide pairs). We denote the complement of some $X\in \Sigma$ by $\overline{X}$.
For some  mRNA with nucleotide sequence $S$,
we define the structure graph of $S$ by taking the nucleotides as
vertices and edges between any two vertices representing complementary nucleotides.

That is, to solve the MRSO problem we have to compute an admissible labeling over $\Sigma$ (i.e. a labeling that satisfies the complementary conditions) for the vertices of the given structure graph of highest possible value with respect to functions $f_i$, $i=1,\ldots,n$.

\begin{problem}[MRSO]
~ \newline
INSTANCE: A structure graph $G=(\{v_1,\ldots, v_{3n}\},E_G)$, and $n$ functions
$f_1,\ldots,f_n$, $f_i:\Sigma^3 \to \mathbb Q$ is associated with $\{v_{3i-2},v_{3i-1},v_{3i}\}$,
 $1\leq i \leq n$.
 \newline
OUTPUT: A function $L: V_G \to \Sigma$, such that $\{v_k,v_l\}\in E_G$ implies that $(L(v_k),L(v_l))\in\Gamma$ and the cost \[MRSO(G,f_1,\ldots,f_n):=\sum_{i=1}^{n} f_i(L(v_{3i-2}),L(v_{3i-1}),L(v_{3i}))\] is maximized.
\end{problem}

In several motivations from biology, the structure graph of problem
MRSO has vertex degree at most one. Following the notions of \cite{Bon04},  we denote the corresponding problem by {\em MRSO-d1}.

\begin{figure}[ht]
\begin{center}
\begin{picture}(245.5,85)
{
\put(10,25){$G$}
\put(60,0){\epsfig{figure=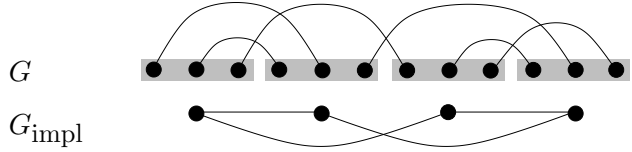,width=6.5cm}}
\put(10,5){$G_{\mbox{\small impl}}$}}
\end{picture}
\end{center}
\caption{A structure graph $G$ and the corresponding implied structure graph $G_{\mbox{\small impl}}$. Vertices corresponding to one codon are framed in a grey box.}
\label{Fgr2}
\end{figure}

Since functions $f_i$, $i=1,\ldots,n$ correspond to $n$ amino acids, we next
describe the MRSO problem on the amino acid level instead
of the given nucleotide level definition. For some structure graph $G=(\{v_1,\ldots, v_{3n}\},E_G)$ we define  the {\em implied structure graph} $G_{\mbox{\small impl}}=(V_{\mbox{\small impl}},E_{\mbox{\small impl}})$  by
$V_{G_{\mbox{\small impl}}}     = \{u_1,\ldots,u_n\}$ and
$E_{G_{\mbox{\small impl}}}    = \{\{u_i,u_j\}~|~ \exists r\in \{3i-2,3i-1,3i\} :   \exists s\in \{3j-2,3j-1,3j\}: \{v_r,v_s\} \in E_G$\}. 
Fig. \ref{Fgr2} shows an example for a structure graph and the corresponding implied structure graph. 

Next we  generalize the complementary conditions given by $\Gamma$ 
for amino acids, i.e. strings of $\Sigma^3$. 
Let $(l_{3i-2}l_{3i-1}l_{3i},l_{3j-2}l_{3j-1}l_{3j})\in \Sigma^3\times \Sigma^3$ be a pair and $v_{3i-2},v_{3i-1},v_{3i}$, $v_{3j-2},v_{3j-1},v_{3j}$  be the six corresponding vertices of structure graph $G$. 
We define pair $(l_{3i-2}l_{3i-1}l_{3i},l_{3j-2}l_{3j-1}l_{3j})$  {\em satisfies $\Gamma$}, if for every edge $\{v_{i'},v_{j'}\}\in E_G$, $3i-2\leq i'\leq 3i$, $3j-2\leq j'\leq 3j$: $(l_{i'},l_{j'})\in \Gamma$.

Obviously, every solution for the MRSO problem on a structure graph $G$ can be
transformed into a solution for the corresponding implied structure graph $G_{\mbox{\small impl}}$, and vice versa. Further for every structure graph $G$ which is an instance 
of problem MRSO-d1, every vertex in $G_{\mbox{\small impl}}$ 
has at most 3 adjacent edges as shown in the example of Fig \ref{Fgr2}.

The following results for the MRSO-d1 problem have been shown.
Problem MRSO-d1 is known to be NP-complete for implied structure graphs with page number at most 2, see \cite{BFHV05}, and thus
for planar implied structure graphs, further in \cite{Bon04} it is shown that
MRSO-d1 generalizes the Maximum independent set problem for graphs of vertex degree at most 3, which is also known to be NP-complete \cite{GJ79}.
Even the decision problem, where an input graph $G$ and $n$ functions $f_1,\ldots,f_n$ are
accepted if some assignment of the vertices reach costs of $c$ is NP-complete for implied structure graphs of vertex degree
at most 3 \cite{BNS02}.

If the implied structure graph is outer-planar, MRSO-d1 is solvable in linear time \cite{BNS02}. Further in \cite{BFHV05} polynomial  fixed parameter algorithms for MRSO-d1 for implied structure graphs of a bounded number of edge crossings, implied structure graphs of a bounded number of degree 3 vertices, and implied structure graphs of a bounded cut-width
(which also implies a polynomial solution for MRSO-d1 on tree-width bounded graphs)
are given.

Since there also exist mRNA structures with bonds between more than two nucleotides  \cite{Aku00} and  amino acids of more complicated secondary structures \cite{BFHV05},
we next give a more general solution for problem MRSO for implied structure graphs of bounded clique-width.

\section{MRSO on implied structure graphs of bounded clique-width}

We next will use the scheme of Theorem \ref{T1} to obtain a polynomial
time solution for the MRSO
problem for associated implied structure graphs of bounded clique-width.

\begin{theorem}
For every positive integer $k$, problem MRSO can be solved in polynomial time for every structure graph that defines an implied structure graph which is given by some clique-width $k$-expression.
\end{theorem}

\begin{proof} Let $G$ be a structure graph for the implied structure graph
$G_{\mbox{\small impl}}=(\{u_1,\ldots,u_n\},E_G,$ $\lab_G)$, which is defined by some clique-width $k$-expression $X$. 
For every admissible labeling $\lab^{\Sigma}:V_G\to\Sigma^3$ of $\val(X)$ we define a pair $(L,f)$, 
where $L=\{(\lab_{\graph(X)}(u),\lab^{\Sigma}(u))~|~ u \in V_{\graph(X)}\}  \subseteq [k]\times \Sigma^3$   and $f=\sum_{u_i\in \val(X)} f_i(\lab^{\Sigma}(u_i))$.
Let $F(X)$ be the set of all mutually different pairs $(L,f)$ for all
admissible labelings of the vertices of graph $\graph(X)$ with labels of $\Sigma$.
Then $F(X)$ is polynomially bounded in the size of $X$, because
$F(X)$ has at most
$(|V|-1)^{|\Sigma|^3\cdot k} \cdot |V|^{|\Sigma|^3}$
mutually different pairs. Each pair contains a label set with
at most $|\Sigma|^3\cdot k$ different pairs of $[k]\times \Sigma^3$ and a sum of at most $|\Sigma|^3$ different
addends.

The following observations show that for every fixed integer $k$, $F(\bullet_a)$, $a\in[k]$, is computable in time $O(1)$, $F(X \oplus Y)$ is computable
in polynomial time from $F(X)$ and $F(Y)$, and $F(\eta_{a,b}(X))$ and $F(\rho_{a \to b}(X))$, $a,b\in[k]$,
are computable in polynomial time from $F(X)$.

\begin{enumerate}
\item
If $\val(X)$ consists of a single vertex $u_i$, then

$F(\bullet_a)=\{(\{(a,l)\},f_i(l))~|~ l\subseteq \Sigma^3 \}$

\item
$F(X \oplus Y)$ is the set of all pairs $(L\cup L',f+f')$ which can be obtained by
a pair $(L,f)\in F(X)$ and a pair  $(L',f')\in F(Y)$.

\item
$F(\eta_{a,b}(X)) = \{ (L,f)\in F(X)
\mid ~ (a,l_1),(b,l_2)\in L \Rightarrow (l_1,l_2) {~ \rm satiesfies ~} \Gamma\}$

\item
$F(\rho_{a \to b}(X)) =\{ (\{(\rho_{a \to b}(a_1),l_1),\ldots,(\rho_{a \to b}(a_m),l_m$ $)\},f) ~|~(\{(a_1,l_1),\ldots,(a_m,l_m)\},f)\in F(X) \}$
\end{enumerate}

There is an admissible labeling of the vertices of $\graph(X)$ with cost $f$ if and only if there is some pair $(L,f) \in F(X)$. The corresponding labeling of the vertices of $\graph(X)$ from $\Sigma^3$ can be
recomputed from expression $X$. By Theorem \ref{T1} the results follows. $\Box$
\end{proof} 

By Theorem \ref{T2} we conclude our main result of this section.

\begin{theorem}
MRSO is computable in polynomial time for every class of structure graphs that define implied structure graphs of bounded clique-width.
\end{theorem}

Since every class of graphs of bounded tree-width has bounded clique-width \cite{CR05}, our result implies that even the MRSO problem can be solved in polynomial time for
structure graphs which define implied structure graphs of bounded tree-width 
which has  been shown in \cite{BFHV05} for the MRSO-d1 problem.

Note that our solution is independent of alphabet $\Sigma$ and set of complementary pairs $\Gamma$, it is only important that $\Sigma$ has a bounded size.

\section{Comparing two solutions of MRSO-d1}

In this section we consider for two given implied structure graphs $G_1$, $G_2$, and two sequences of similarly functions $f_i$, $g_i$, $1\leq i \leq n$, the complexity of comparing the costs of the corresponding two solutions of problem MRSO-d1. We will show that these
compare problems are even complete for the complexity class $\pnp$, which is 
assumed to be a strong super set of NP. Class $\pnp$ is defined as the set of
problems which can be solved in polynomial time with a number of parallel
queries to an oracle in NP. For more
results concerning $\pnp$-hardness see \cite{Wag87,SV00}.

We next assume the restricted case that $\Sigma=\{a,b,\overline{a},\overline{b}\}$ and  $\Gamma=\{(a,\overline{a}),(b,\overline{b})\}$, see \cite{Bon04}.

\begin{problem} [Comparing (Equality) MRSO-d1] \label{pa}
~ \newline
INSTANCE: Two structure graphs $G_1=(\{v_{1},\ldots,v_{3n}\},E_1)$ and $G_2=(\{u_{1},\ldots,$ $u_{3m}\},E_2)$, $n$ functions
$f_1,\ldots,f_n$, $f_i:\Sigma^3 \to \mathbb Q$,  and $m$ functions
$g_1,\ldots,g_m$, $g_i:\Sigma^3 \to \mathbb Q$.

QUESTION: Is MRSO-d1$(G_1,f_1,\ldots,f_n) \leq$ MRSO-d1$(G_2,g_1,\ldots,g_m)$?

(QUESTION: Is MRSO-d1$(G_1,f_1,\ldots,f_n) =$ MRSO-d1$(G_2,g_1,\ldots,g_m)$?)
\end{problem}

\begin{theorem}
Comparing MRSO-d1 and Equality MRSO-d1 is $\pnp$-complete for planar graphs.
\end{theorem}

\begin{proof}
First we have to show that Comparing MRSO-d1 and Equality MRSO-d1  is contained in $\pnp$.
Therefor, we define a polynomial time algorithm solving the problem Comparing MRSO-d1 with a number of parallel queries to an oracle in NP.
We take the MRSO-d1 problem as our oracle, which is in NP. 
Given two graphs $G_1$ and $G_2$, we ask the oracle  the following two queries:
MRSO-d1$(G_1,f_1,\ldots,f_n)$
and 
MRSO-d1$(G_2,g_1,\ldots,g_n)$.
We accept for the problem Comparing MRSO-d1 if both values are equal. Analogously
we can define  parallel queries to an oracle in NP for the problem 
Equality MRSO-d1.

In \cite{SV00} the problem of comparing the maximum vertex cover of two graphs
has been shown to be $\pnp$-complete. Using the reductions  of \cite{GJS76a} Theorem 2.7  and \cite{GJ77} Lemma 1 we conclude that comparing the maximum vertex cover of two graphs remains $\pnp$-complete for planar graphs of vertex degree at most 3.

Since every vertex cover $C\subseteq V$ of a graph 
$G=(V,E)$, obviously  corresponds to an independent set $V-C$ in graph $G$, 
we conclude that comparing the maximum independent set of two graphs remains $\pnp$-complete for planar graphs of vertex degree at most 3.

This allows us to 
show the $\pnp$-hardness of Comparing MRSO-d1 and Equality MRSO-d1 by a reduction from comparing independent set for planar graphs of vertex degree at most 3 by the idea of the proof of  Theorem 3 in \cite{Bon04}. 
Given an instance $G$ of maximum independent set of vertex degree at most 3,
the proof constructs an instance $(G_{\mbox{\small impl}},f_1,\ldots,f_n)$ for problem MRSO-d1, such that MRSO-d1$(G_{\mbox{\small impl}},f_1,\ldots,f_n)$ is equal to the maximum independent set of $G$.
$\Box$
\end{proof}

\bibliographystyle{alpha}
\bibliography{/home/gurski/bib.bib}
\end{document}